\documentclass[reprint,twocolumn,pre,showpacs,amsmath,amssymb,aps,superscriptaddress]{revtex4-1}

\usepackage{natbib}
\usepackage{graphicx,color,rotating}
\usepackage[latin1]{inputenc}
\usepackage{textcomp}
\usepackage{dcolumn}            
\usepackage{bm}                 
\usepackage{mathtools}          
\usepackage{upgreek}
\usepackage{subdepth}  			
\usepackage{epstopdf}   		
\usepackage{float}				

\usepackage{array}
\newcolumntype{L}[1]{>{\raggedright\let\newline\\\arraybackslash\hspace{0pt}}m{#1}}
\newcolumntype{C}[1]{>{\centering\let\newline\\\arraybackslash\hspace{0pt}}m{#1}}
\newcolumntype{R}[1]{>{\raggedleft\let\newline\\\arraybackslash\hspace{0pt}}m{#1}}

\usepackage{hyperref}					
\usepackage{xcolor}						
\hypersetup{			     			
    colorlinks,
    allcolors={blue}
}
\newcommand{\etal}{\textit{et~al.}}

\newcommand{\qmarks}[1]{{``#1''}}
\newcommand{\upspace}{\rule{0ex}{2.5ex}}
\newcommand{\downspace}{\rule[-1.5ex]{0ex}{1.0ex}}

\newcommand{\mr}[1]{\ensuremath{\mathrm{#1}}}
\newcommand{\myvec}[1]{\bm{#1}}
\newcommand{\ee}{\mathrm{e}}
\newcommand{\ii}{\mathrm{i}}
\newcommand{\dm}{\mathrm{d}}

\DeclareMathOperator{\re}{Re}

\newcommand{\iot}{{\ii\omega t}}

\newcommand{\ve}{\varepsilon}

\newcommand{\pp}{\partial}

\newcommand{\nablabf}{\boldsymbol{\nabla}}

\newcommand{\DDD}{\myvec{D}}

\newcommand{\eee}{\myvec{e}}

\newcommand{\JJJ}{\myvec{J}}

\newcommand{\nnn}{\myvec{n}}

\newcommand{\rrr}{\myvec{r}}

\newcommand{\uuu}{\myvec{u}}

\newcommand{\zerovec}{\boldsymbol{0}}

\newcommand{\Gamsl}{\Gamma_\mathrm{sl}}

\newcommand{\SIC}{\textrm{C}}
\newcommand{\SICel}{^\circ\!\textrm{C}}

\newcommand{\SIGHz}{\textrm{GHz}}
\newcommand{\SIMHz}{\textrm{MHz}}

\newcommand{\SIkg}{\textrm{kg}}

\newcommand{\SIm}{\textrm{m}}

\newcommand{\SImum}{\textrm{\textmu{}m}}

\newcommand{\SIGPa}{\textrm{GPa}}

\newcommand{\SIs}{\textrm{s}}

\newcommand{\SImps}{\SIm\,\SIs^{-1}}

\newcommand{\beq}[1]{\begin{equation} \eqlab{#1}}
\newcommand{\eeq}{\end{equation}}
\newcommand{\bsub}{\begin{subequations}}
\newcommand{\esub}{\end{subequations}}
\def\bal#1\eal{\begin{align}#1\end{align}}
\def\bsubal#1\esubal{\bsub \begin{align}#1\end{align} \esub}
\newcommand{\nn}{\nonumber}
\newcommand{\eqlab}[1]{\label{eq:#1}}
\renewcommand{\eqref}[1]{Eq.~(\ref{eq:#1})}
\newcommand{\eqnoref}[1]{(\ref{eq:#1})}

\newcommand{\eqsnoref}[2]{(\ref{eq:#1}) and~(\ref{eq:#2})}

\newcommand{\figref}[1]{Fig.~\ref{fig:#1}}

\newcommand{\figlab}[1]{\label{fig:#1}}

\newcommand{\secref}[1]{Section~\ref{sec:#1}}

\newcommand{\seclab}[1]{\label{sec:#1}}
\newcommand{\tabref}[1]{Table~\ref{tab:#1}}

\newcommand{\tablab}[1]{\label{tab:#1}}

\newcommand{\sigmabf}{\bm{\sigma}}

\newcommand{\cL}{c_\mathrm{lo}}
\newcommand{\cT}{c_\mathrm{tr}}

\begin{document}

\title{Fabrication, characterization, and simulation of glass devices with AlN-thin-film-transducers for excitation of ultrasound resonances}

\author{Andr\'e G. Steckel}
\email{angust@fysik.dtu.dk}
\affiliation{Department of Physics, Technical University of Denmark,
DTU Physics Building 309, DK-2800 Kongens Lyngby, Denmark}

\author{Henrik Bruus}
\email{bruus@fysik.dtu.dk}
\affiliation{Department of Physics, Technical University of Denmark,
DTU Physics Building 309, DK-2800 Kongens Lyngby, Denmark}

\author{Paul Muralt}
\email{paul.muralt@epfl.ch}
\affiliation{Materials Science, Station 12, EPFL,  1015 Lausanne, Switzerland}
\affiliation{PIEMACS S\`arl, EPFL Innovation Parc,
B\^atiment C, 1015-Lausanne Switzerland}

\author{Ramin Matloub}
\email{ramin.matloub@piemacs.ch}
\affiliation{PIEMACS S\`arl, EPFL Innovation Parc,
B\^atiment C, 1015-Lausanne Switzerland}

\date{16 November 2020}

\begin{abstract}
We present fabrication of 570-$\SImum$-thick, millimetric-sized soda-lime-silicate float glass blocks with a 1-$\SImum$-thick AlN-thin-film piezoelectric transducer sandwiched between thin metallic electrodes and deposited on the top surface. The electromechanical properties are characterized by electrical impedance measurements in the frequency range from 0.1 to 10~MHz with a peak-to-peak voltage of 0.5~V applied to the electrodes. We measured the electrical impedance spectra of 35 devices, all of width 2~mm, but with 9 different lengths ranging from 2 to 6~mm and with 2-7 copies of each individual geometry. Each impedance spectrum exhibits many resonance peaks, of which we carefully measured the 5 most prominent ones in each spectrum. We compare the resulting 173 experimental resonance frequencies with the simulation result of a finite-element-method model that we have developed. When using material parameters from the manufacturer, we obtain an average relative deviation of the 173 simulated resonance frequencies from the experimental ones of $(-4.2\pm0.04)$\%. When optimizing the values of the Young's modulus and the Poisson ratio of the float glass in the simulation, this relative deviation decreased to $(-0.03\pm0.04)$\%. Our results suggest a method for an accurate \emph{in-situ} determination of the acoustic parameters at ultrasound frequencies of any elastic solid onto which a thin-film transducer can be attached.
\end{abstract}


\maketitle

\section{Introduction}
\seclab{intro}

Aluminum nitride (AlN) is one of the most commonly used materials in integrated thin-film piezoelectric transducers for actuating electromechanical systems (MEMS). Due to their low dielectric and mechanical loss tangent, their structural and chemical stability, as well as their compatibility with standard silicon-based CMOS microfabrication techniques, AlN-sputtered thin films are commercially used in thin film bulk-wave acoustic resonator filters~\cite{Ruby2017}. The academic literature reports applications of AlN thin-film transducers as RF filters~\cite{Dubois1999}, contour mode resonators~\cite{Piazza2006}, switches~\cite{Zaghloul2014, Sinha2009}, suspended microchannel resonators~\cite{DePastina2018}, and accelerometers~\cite{Olsson2009}. Detailed studies on AlN-thin-film-actuated high-tone bulk acoustic resonators with high quality factors have been studied in the 0.3 - 3~GHz range on 350-$\SImum$-thick substrates of sapphire, crystal quartz, fused silica, and silicon \cite{Zhang2006}, and on 30-$\SImum$-thick Si membranes \cite{Masson2007, Fujikura2008}.

Because AlN is a nonferroelectric polar material, a reorientation of the polar axis is not possible. The material growth process therefore has to provide for a textured structure that includes the alignment of the polar directions~\cite{Muralt2008}, since otherwise the random orientation of the grains constituting the AlN thin film would result in a zero average piezoelectric effect. In this work, we therefore follow the process developed in our previous work on pure~\cite{Howell2019} and Sc-doped~\cite{Matloub2011, Matloub2013} AlN films, and select Pt/Ti as the seed layer for the bottom electrode stack as it promotes nucleation of (002)-oriented AlN grains.

The past few years, piezoelectric thin films made of nonferroelectric polar materials, such as zinc oxide (ZnO) and AlN, have found a broad range of lab-on-chip applications such as biosensing, particle/cell concentrating, sorting, patterning, pumping, mixing, nebulization and jetting. Integrated devices with acoustic transducers, sensors, and microfluidic channels have been fabricated by depositing these piezoelectric films onto wide range of materials beyond the usual silicon substrates such as ceramics, diamond, glass, and more recently also polymer, metallic foils and flexible glass for making flexible devices. Such thin-film acoustic wave devices have great potential for implementing integrated, disposable, or bendable lab-on-a-chip devices into various sensing and actuating applications~\cite{Fu2017}. To advance such applications, we focus in this work on AlN-thin-film transducers deposited on glass substrates and driven in the 0.1 - 10~MHz range.

The deposition of AlN thin films and their texture control are more difficult than in the case with ZnO films~\cite{Lee2007}, and their integration into MEMS is not a trivial task, as they tend to have high in-plane stress with a sharp transition around zero stress~\cite{Dubois2001}. However, the thermal conductivity of AlN is more than 3 times larger than of ZnO, which enables the operation of AlN films at a higher power level, an important feature for applications in RF technology. AlN films are normally sputter-deposited at moderate temperatures (typically 200-400~$\SICel$) to achieve an optimal performance, although \qmarks{room temperature} deposition with a local temperature near the target of around 150~$\SICel$ is sometimes used~\cite{Lee2002}. Also deposition conditions, especially oxygen or moisture in vacuum chambers, have significant effects on AlN film growth and microstructure. Growing AlN films thicker than a few microns is particularly challenging because of the potentially large film stress and the tendency to form microcracks.

In this work, we used our titanium-seed-layer technique~\cite{Howell2019} to improve the fabrication of AlN-thin-film transducers deposited on glass substrates for future use in biosensing and lab-on-a-chip applications. In \secref{design}, we describe the design of the devices, followed in \secref{fabrication} by a presentation of the fabrication and characterization processes.  In \secref{Zexp}, we then characterize the acoustic properties of the devices by measuring electrical impedance spectra in the frequency range from 0.1 to 10~MHz. The numerical model and its finite-element-method implementation is described in \secref{modeling}. The comparison of the simulation results with the experiments presented in \secref{results} are in so good agreement that our numerical handling of the measured impedance spectra can be turned into an improved \emph{in-situ} determination of the acoustic properties of the glass substrate. Finally, concluding remarks are presented in \secref{conclusion}.

\begin{figure}[!b]
\centering
\includegraphics[width=0.95\columnwidth]{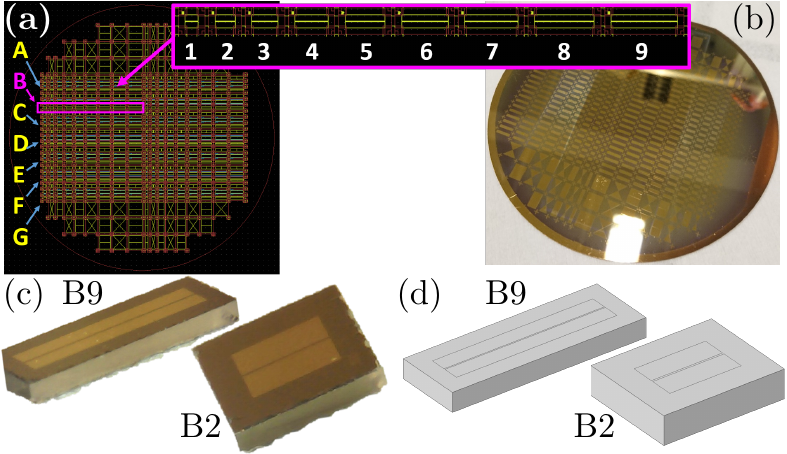}
\caption{\figlab{DesignFabrication} (a) Design of the electrode mask for the lithography step in the fabrication process that defines the gold electrodes on the top surface of the glass wafer. The relevant devices are taken from row A - G and column 1 - 9. The inset show row B column 1 - 9. Devices with the same column index are nominally identical. (b) A photograph of the fabricated wafer before dicing it in individual devices.  (c) Microscope images of the actual device B2 and B9. (d)  Rendering of the two 2-mm-wide and 0.58-mm-high devices B2 (2.5 mm long) and B9 (6 mm long), showing the outline of the 1-mm-wide top electrode split in two halves by a 40-$\SImum$-wide gap.}
\end{figure}

\section{Device design}
\seclab{design}

In \figref{DesignFabrication} is shown the design of the devices under study. This design is chosen to ease the microfabrication, the experimental characterization, and the subsequent validation of a finite-element method model of the system. All devices consist of a simple box-shaped 570-$\SImum$-thick and 2-mm-wide block of soda-lime-silicate (SLS) float glass~\cite{Schott_float_glass} of length $L$ diced from a wafer.  $L$ is in the range from 2 to 6~mm in steps of 0.5~mm, thus resulting in 9 groups of nominally identical devices. A 1-$\SImum$-thick AlN-thin-film piezoelectric transducer, sandwiched between sub-$\SImum$-thick metal electrodes, is deposited on top surface of the glass block. To ease the dicing process and to avoid short circuiting between the two electrodes, the top electrode of the transducer has its width reduced to 1~mm and its length to $L_\mr{te} = L - 1$~mm. Moreover, to allow for anti-symmetric voltage actuation, the top electrode is split in half by a 40-$\SImum$-wide gap perpendicular to the length direction.  As listed in \tabref{device_list}, there are up to seven nominally identical copies A, B, C, D, E, F, and G in each group of devices.

\begin{table}[t]
\centering
\vspace*{-2.5mm}
\caption{\tablab{device_list} A list of the 35 devices under study. The rectangular glass blocks are all 0.58~mm thick, 2~mm wide, and with a length $L$ in the range from 2 to 6~mm. The deposited AlN-thin-film piezoelectric transducer is 1~$\SImum$ thick and sandwiched between 0.1-$\SImum$-thick metal electrodes. The top electrode has the width 1~mm and length $L_\mr{te} = L - 1$~mm. All devices with the same column index are nominally identical.} \begin{ruledtabular}
 \begin{tabular}{clcc}
 Device & Device & $L$  & $L_\mr{te}$ \\
 column  & row  & [mm] & [mm]  \downspace\\ \hline
 1 & C, D, G & 2.0 & 1.0 \\
 2 & C, D, E, F, G & 2.5 & 1.5 \\
 3 & A, B, C, G & 3.0 & 2.0 \\
 4 & A, B, D & 3.5 & 2.5 \\
 5 & A, B, C & 4.0 & 3.0 \\
 6 & B, C & 4.5 & 3.5 \\
 7 & A, B, C, D, E & 5.0 & 4.0 \\
 8 & A, B, C, D, E, F, G & 5.5 & 4.5 \\
 9 & C, D, E& 6.0 & 5.0 \\
 \end{tabular}
\end{ruledtabular}
\end{table}


\section{Fabrication and characterization of the piezoelectric thin-film transducer}
\seclab{fabrication}

The device fabrication is based on a 4-inch SLS float glass wafer (SCHOTT Suisse SA, Yverdon). This glass has a low thermal expansion, a high thermal shock resistance, and an ability to withstand temperatures up to $490~\SICel$ for long periods, so by not raising the process temperatures above $300~\SICel$, we ensure a high quality and reproducibility in our fabrication process. Following a microfabrication process similar to the one described in Ref.~\cite{Howell2019} and sketched in \figref{fabrication}, the first step is cleaning of the glass substrate with isopropyl alcohol and acetone to remove organic contaminations. Additional plasma cleaning was performed right before the deposition of the bottom electrode in the deposition tool. We prepared textured Ti, Pt, AlN thin films using a dedicated cluster sputtering chamber (Pfeiffer Vacuum), and then sputter-deposited 10~nm Ti at $300~\SICel$ serving as an adhesion layer between the glass substrate and the Pt bottom electrode. Next, the 100-nm thick Pt electrode was deposited with the same tool in a different chamber also at $300~\SICel$ and without breaking the vacuum. Successively, $1~\SImum$ (0001)-textured AlN was deposited using a dedicated reactive, pulsed, direct current magnetron sputtering chamber at $300~\SICel$, and on top of the AlN film 10~nm Cr and 150~nm Au were sputter deposited as top electrode with Cr being the adhesion layer between the AlN and Au. As the next process step, standard photolithography was exploited to transfer the intended pattern on to the top electrode, and this pattern was realized by ion-beam etching (Veeco). Finally, the wafer was diced into the smaller chips ready for use, as shown in \figref{DesignFabrication}(c).

After fabrication and dicing, the quality of the resulting devices was checked by visual inspection in a microscope, and their dimensions were checked by measurements using a digital caliper (Vernier Caliper, Cocraft) with 0.03-mm accuracy and 0.01-mm repeatability. 29 representative devices were selected. Their lengths were found to range from  $(2034 \pm 5)~\SImum$ to $(6034 \pm 5)~\SImum$ in steps of 500~$\SImum$, the widths were $(2025 \pm 5)~\SImum$, and the heights were $(570 \pm 3)~\SImum$. The uncertainties represent one standard deviation of several repeated measurements, and the relative uncertainty ranges from 0.6\% for the height, down to 0.1\% for the longest length.

\begin{figure}[!t]
\centering
\includegraphics[width=0.8\columnwidth]{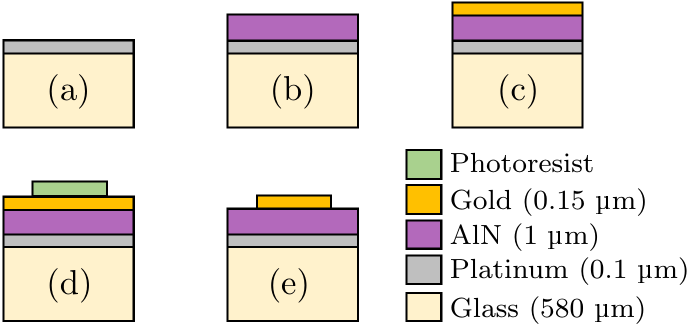}
\caption{\figlab{fabrication} Sketch of the fabrication process, where a stack consisting of a bottom electrode, the AlN thin film, and the top electrode is deposited on the glass block. (a) Deposition of a 10-nm-thick Ti adhesion layer (not shown) followed by a 100-nm-thick platinum layer. (b) Deposition of the 1-$\SImum$-thick AlN film. (c) Deposition of a 10-nm-thick Cr adhesion layer (not shown) followed by a 150-nm-thick gold layer. (d) Deposition and development of photoresist. (e) Patterning of the top electrode by ion-beam etching of the exposed gold followed by removal of the photoresist.}
\end{figure}

\section{Measured electrical impedance}
\seclab{Zexp}
The electrical impedance of the 35 devices listed in \tabref{device_list} were measured as a function of frequency from from 0.1 to 3~MHz in steps of 500~Hz using an Agilent 4294A Precision Impedance Analyzer (Agilent Technologies AG, Basel, Switzerland) with a 0.5-V$_\mr{pp}$ ac-voltage applied in anti-phase between the two halves of the split top electrode. The bottom electrode was left floating.

\begin{figure}[!b]
\centering
\includegraphics[width=1\columnwidth]{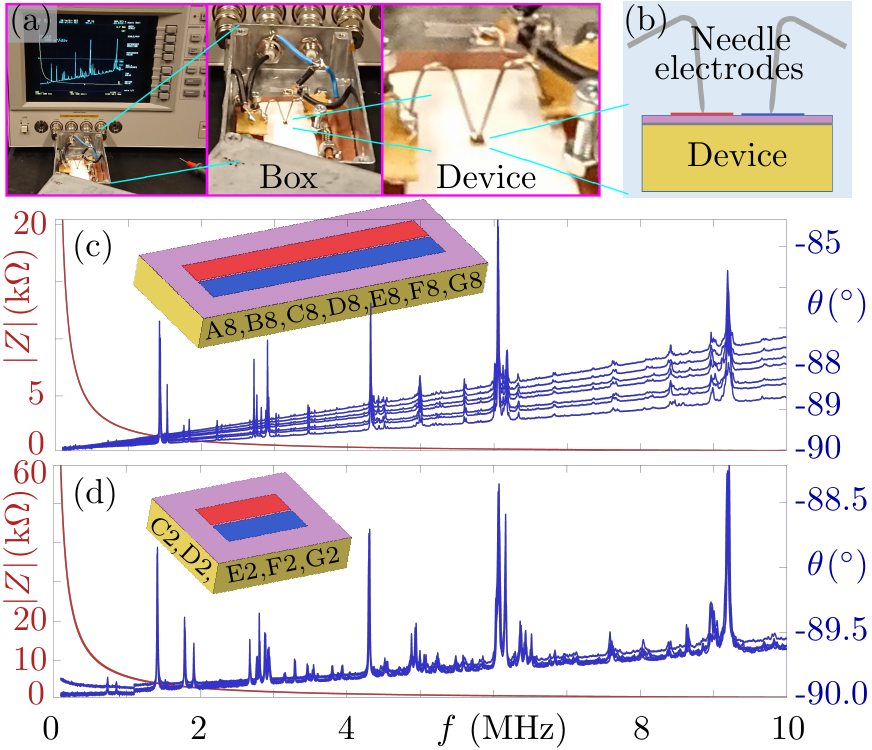}
\caption{\figlab{Setup_and_measurements} (a) Three photographs of the spectrum analyzer and the attached metal box that provides electrical shielding of the device, but here with its lid removed to reveal the device mounted inside.  A zoom-in on the interior of the metal box shows the coax connectors and the wires leading to the device. A further zoom-in shows the spring-loaded electrodes attached to the device. (b) An end-view sketch of the spring-loaded electrodes attached to the device consisting of the glass block (beige) and the AlN thin-film transducer (purple) with a split top electrode (red-blue). (c) The impedance spectra showing the magnitude $|Z(f)|$ (red curves) and the phase $\theta(f)$ (blue curves) for the seven nominally identical devices A8, B8, C8, D8, E8, F8, and G8 shown in the inset. The phase curves differ as the devices were not discharged initially. (d) Same as for panel (c) but for the five  nominally identical devices C2, D2, E2, F2, and G2. Here, the phase curves are nearly identical as the devices were discharged initially.}
\end{figure}

\begin{table*}[t]
\centering
\vspace*{-2.5mm}
\caption{\tablab{FreqResultsExpSim} For each of the nine device geometries, labeled by column index $k = 1$ - 9 in \tabref{device_list}, five resonance peaks $(k,n)$  of increasing frequency with $n = 1$ - 5 were selected  the impedance spectrum for comparing simulations with measurements. Peak $(4,1)$ and $(4,2)$ marked by an asterisk were not measured on device B in column $n = 4$. The experimental value $f^\mr{exp}_{k,n}$ is the average resonance frequency of peak $n$ over all devices (row index A - G in \tabref{device_list}) with column index $k$. The simulation result $f^\mr{sim,A}_{k,n}$ is the corresponding resonance frequency obtained by the numerical modeling described in \secref{modeling} and using the parameter values given in the literature. The simulation result $f^\mr{sim,B}_{k,n}$ is obtained by using the optimized values listed in \tabref{glass_param_fitted} of Young's modulus and Poisson's ratio for the glass block. Finally, for X~=~A or B, $\Delta^\mr{X}_{k,n} = (f^\mr{sim,X}_{k,n} - f^\mr{exp}_{k,n})/f^\mr{exp}_{k,n}$ is the relative deviation of the simulation result from the experimental value for resonance peak $n$ in device geometry $k$.}
\begin{ruledtabular}  %
\begin{tabular}{lddddd@{\qquad\qquad\qquad}cddddd}
Peak &
\multicolumn{1}{r}{\text{$f^\mr{exp}_{k,n}\;$}}    &
\multicolumn{1}{r}{\text{$f^\mr{sim,A}_{k,n}$}}      &
\multicolumn{1}{r}{\text{$f^\mr{sim,B}_{k,n}$}}      &
\multicolumn{1}{l}{\text{$\;\;\Delta^\mr{A}_{k,n}$}} &
\multicolumn{1}{l}{\text{$\;\;\Delta^\mr{B}_{k,n}$}} &
Peak &
\multicolumn{1}{r}{\text{$f^\mr{exp}_{k,n}\;$}}    &
\multicolumn{1}{r}{\text{$f^\mr{sim,A}_{k,n}$}}      &
\multicolumn{1}{r}{\text{$f^\mr{sim,B}_{k,n}$}}      &
\multicolumn{1}{l}{\text{$\;\;\Delta^\mr{A}_{k,n}$}} &
\multicolumn{1}{l}{\text{$\;\;\Delta^\mr{B}_{k,n}$}}
\\
$k,n$&
\multicolumn{1}{r}{[MHz]} &
\multicolumn{1}{r}{[MHz]} &
\multicolumn{1}{r}{[MHz]} &
\multicolumn{1}{l}{\;\;\;[\%] } &
\multicolumn{1}{l}{\;\;\;[\%] } &
$k,n$ &
\multicolumn{1}{r}{[MHz]} &
\multicolumn{1}{r}{[MHz]} &
\multicolumn{1}{r}{[MHz]} &
\multicolumn{1}{l}{\;\;\;[\%] } &
\multicolumn{1}{l}{\;\;\;[\%] } \downspace\\ \hline
1, 1    & 0.943  & 0.883  & 0.926  &-6.3  & -1.7  & 5, 4  & 2.712  & 2.607  & 2.721   &-3.9  &  0.3 \upspace\\
1, 2    & 1.424  & 1.369  & 1.422  &-3.9  & -0.2  & 5, 5  & 2.901  & 2.780  & 2.918   &-4.2  &  0.6 \\
1, 3    & 2.065  & 1.958  & 2.054  &-5.2  & -0.5  & 6, 1  & 1.425  & 1.368  & 1.422   &-4.0  & -0.2 \\
1, 4    & 2.717  & 2.618  & 2.720  &-3.7  &  0.1  & 6, 2  & 1.537  & 1.470  & 1.538   &-4.3  &  0.1 \\
1, 5    & 2.896  & 2.766  & 2.906  &-4.5  &  0.3  & 6, 3  & 1.606  & 1.540  & 1.601   &-4.1  & -0.3 \\
2, 1    & 0.737  & 0.701  & 0.736  &-4.9  & -0.2  & 6, 4  & 2.741  & 2.641  & 2.739   &-3.7  & -0.1 \\
2, 2    & 1.418  & 1.365  & 1.419  &-3.7  &  0.1  & 6, 5  & 2.910  & 2.748  & 2.922   &-5.6  &  0.4 \\
2, 3    & 1.791  & 1.706  & 1.788  &-4.8  & -0.2  & 7, 1  & 0.321  & 0.306  & 0.322   &-4.8  &  0.3 \\
2, 4    & 1.915  & 1.830  & 1.906  &-4.5  & -0.4  & 7, 2  & 1.440  & 1.379  & 1.431   &-4.2  & -0.6 \\
2, 5    & 2.942  & 2.802  & 2.939  &-4.7  & -0.1  & 7, 3  & 1.561  & 1.495  & 1.555   &-4.2  & -0.4 \\
3, 1    & 0.612  & 0.582  & 0.611  &-4.9  & -0.2  & 7, 4  & 2.733  & 2.632  & 2.731   &-3.7  & -0.1 \\
3, 2    & 1.403  & 1.348  & 1.409  &-3.9  &  0.4  & 7, 5  & 2.905  & 2.783  & 2.920   &-4.2  &  0.5 \\
3, 3    & 1.489  & 1.424  & 1.482  &-4.3  & -0.5  & 8, 1  & 1.438  & 1.375  & 1.426   &-4.4  & -0.9 \\
3, 4    & 1.752  & 1.681  & 1.754  &-4.0  &  0.1  & 8, 2  & 1.533  & 1.470  & 1.531   &-4.1  & -0.2 \\
3, 5    & 2.735  & 2.640  & 2.737  &-3.5  &  0.1  & 8, 3  & 1.834  & 1.760  & 1.835   &-4.0  &  0.1 \\
4, 1$^*$& 0.522  & 0.498  & 0.527  &-4.6  &  0.9  & 8, 4  & 2.718  & 2.618  & 2.725   &-3.7  &  0.3 \\
4, 2$^*$& 0.544  & 0.519  & 0.541  &-4.6  & -0.4  & 8, 5  & 2.903  & 2.786  & 2.922   &-4.0  &  0.7 \\
4, 3    & 1.431  & 1.374  & 1.426  &-4.0  & -0.3  & 9, 1  & 1.429  & 1.373  & 1.426   &-3.9  & -0.2 \\
4, 4    & 1.667  & 1.600  & 1.669  &-4.0  &  0.1  & 9, 2  & 1.511  & 1.451  & 1.511   &-4.0  &  0.0 \\
4, 5    & 2.725  & 2.627  & 2.728  &-3.6  &  0.1  & 9, 3  & 1.765  & 1.696  & 1.768   &-3.9  &  0.2 \\
5, 1    & 1.429  & 1.372  & 1.425  &-4.0  & -0.3  & 9, 4  & 2.089  & 2.018  & 2.111   &-3.4  &  1.1 \\
5, 2    & 1.609  & 1.545  & 1.611  &-4.0  &  0.1  & 9, 5  & 2.744  & 2.644  & 2.746   &-3.7  &  0.1 \\
5, 3    & 2.104  & 2.015  & 2.107  &-4.2  &  0.2  &       &        &        &         &      &      \\
\end{tabular}
\end{ruledtabular}

\end{table*}

To reduce the electrical noise, a given device under study was placed in a grounded metal box, from where it was connected by short coax cables to the impedance analyzer by pressing a spring-loaded needle electrodes to each half of the split top electrode, see \figref{Setup_and_measurements}(a)-(b). The setup was also designed to minimize undesired parasitic components: the parasitic inductance in the system was reduced to somewhere around $L_\mr{ext} = (28 \pm 15)$~nH, the parasitic resistance was roughly $R_\mr{ext} = (5 \pm 3)~\mr{\Omega}$, and the parasitic capacitance was negligible compared to the intrinsic capacitance $C_\mr{dev}$ of the device formed by the electrodes separated by the thin AlN film, and ranging from 18~pF for the smallest device to 86~pF for the largest. This design ensured that the resonance frequency $f_\mr{ext} = \frac{1}{2\pi}(L_\mr{ext}C_\mr{dev})^{-\frac12}$ involving the external inductance  was pushed up in value, $100~\SIMHz \lesssim f_\mr{ext} \lesssim 220~\SIMHz$, much larger than 10~MHz, and therefore not interfering with the intrinsic resonances of the device under study. Finally, the device under study was placed on a folded, soft piece of tissue paper to acoustically isolate from the environment.

Examples of the recorded 35 impedance spectra $Z(f) = |Z|\:\ee^{\ii \theta}$ are shown in \figref{Setup_and_measurements}(c) and (d), where we plot the magnitude $|Z(f)|$ and the phase $\theta(f)$. We first note that the intrinsic resonances of the devices are most clearly visible in the phase plots $\theta(f)$. Due to the parasitic impedances of the system, the base level of $\theta(f)$ is a linear increase from the $-90^\circ$ of an ideal capacitor in the dc limit to a value between $-89.7^\circ$ and  $-87.8^\circ$ at 10~MHz. The device resonances appear as sharp peaks in the phase-curve with a width of around 0.01 times the resonance frequency itself and with an amplitude ranging from 0.1$^\circ$ to 3.5$^\circ$.
This technique is a variant of standard resonant ultrasound spectroscopy (RUS) \cite{Radovic2004, Cachiaras2017}. However, in contrast to conventional RUS, the ultrasound transducers in our case are an integrated but tiny part of the structures to be characterized.

Next, we note that position and amplitudes of the resonance peaks are nearly identical for samples with nominally identical geometry from different parts of the wafer, see \figref{DesignFabrication}(a), for the about 30 peaks easily identified in the range from 0.1 to 10~MHz. About 40\% of the devices selected from column 1 - 9 and row A - G had a short circuit between one or both of the top electrodes with the bottom electrode, so they were discarded from the study. This is the reason that only column~8 in \tabref{device_list} all seven fabricated devices. We found that the slope of the base level of the phase spectra $\theta(f)$ could vary even for nominally identical devices, as seen in \figref{Setup_and_measurements}(d). It turned out that this variance was due to spurious charging effects, perhaps due to the floating electrode. This effect could be avoided by shortly short-circuiting the electrodes with a piece of metallic wire, and thus discharging the system. After discharging, the phase spectra became nearly identical for all devices with the same geometry, see \figref{Setup_and_measurements}(e). When taking the mentioned precautions, the impedance spectra exhibits a large degree of repeatability and reproducibility.

For each of the nine different device geometries, labeled by column index $k = 1$ - 9 in \tabref{device_list}, among our 35 selected devices, we measure the resonance frequency of the five most prominent resonance peaks, labeled by $n = 1$ - 5 in the impedance-phase spectra $\theta(f)$. This resulted in $5\times35 - 2 = 173$ resonance frequencies (for device 4B we did not measure all five peaks, but only the three highest). In \tabref{FreqResultsExpSim} we list the experimentally determined values of the resonance frequencies, but to avoid showing all 173 of them, we only show the average resonance frequency $f^\mr{exp}_{k,n}$ of peak $n$ over all devices (row index A - G in \tabref{device_list}) with column index $k$, which results in $9\times 5 = 45$ entries.

\section{Modeling and numerical simulation}
\seclab{modeling}

We want to compare the experimental results for the impedance spectra and resonance frequencies with numerical simulation. We therefore use the three-dimensional model for acoustofluidic systems recently developed by Skov \etal\ \cite{Skov2019}. The model couples the mechanical displacement field $\uuu$ and the electric potential $\varphi$ in a system driven by a time-harmonic electric potential, $\tilde{\varphi} = \varphi_0\:\ee^{-\iot}$ with angular frequency $\omega = 2\pi f$ and frequency $f$, applied to the interfaces between the metal electrodes and the piezoelectric AlN-thin-film transducer. This time-harmonic boundary condition excites the time-harmonic fields: the electric potential $\tilde{\varphi}(\rrr,t)$ in the AlN-thin-film transducer and the displacement $\tilde{\uuu}(\rrr,t)$ in all solids of the system,
 \bal
 \tilde{\varphi}(\rrr,t)=\varphi(\rrr)\:\ee^{-\iot},
 \qquad
 \tilde{\uuu}(\rrr,t)=\uuu(\rrr)\:\ee^{-\iot}.
 \eal
Fields with a tilde are time-harmonic, whereas those without a tilde are complex-valued amplitudes.

The motion of an elastic solid with a given density $\rho$ (in the model it is the glass, the metal electrodes, and the AlN-thin-film transducer) is described by the displacement field $\uuu$, whereas the electrodynamics is described by the electric displacement field $\DDD$. The governing equations for $\uuu$ and $\DDD$ are Cauchy's elastodynamic equation and Maxwell's quasi-electrostatic equation, $\nablabf \cdot \DDD = 0$,
 \beq{EquMotionSolid}
 -\rho\omega^2(1+\ii\Gamsl)\:\uuu = \nablabf \cdot \sigmabf, \quad
  \nablabf \cdot \big[-(1+\ii\Gamma_\ve)\bm{\ve}\cdot\nablabf\varphi\big] = 0.
 \eeq
Here, $\Gamsl,\Gamma_\ve \ll1$ are weak damping coefficients, $\bm{\ve}$ is the dielectric tensor, and $\sigmabf$ is the stress tensor, which is coupled to $\uuu$ through a stress-strain relation depending on the material-dependent elastic moduli $C_{ik}$.

For the isotropic glass and metal electrodes, the electric field is unimportant, whereas the relation between the stress tensor $\sigma^{{}}_{ik}$ and strain components $\frac12(\pp^{{}}_i u^{{}}_k+\pp^{{}}_k u^{{}}_i)$ is given in the compact Voigt representation as
 \beq{StressStrainSolid}
 \left( \begin{array}{c}
 \!\!\sigma^{{}}_{xx}\!\! \\[1mm]
 \!\!\sigma^{{}}_{yy}\!\! \\[1mm]
 \!\!\sigma^{{}}_{zz}\!\! \\[1mm]\hline
 \!\!\sigma^{{}}_{yz}\!\! \\[1mm]
 \!\!\sigma^{{}}_{xz}\!\! \\[1mm]
 \!\!\sigma^{{}}_{xy}\!\! \end{array} \right)
  \!=\!
 \left( \begin{array}{ccc|ccc}
 \!C^{{}}_{11} & C^{{}}_{12} & C^{{}}_{12} & 0 & \!\!0 & \!\!0 \!\!\\[1mm]
 \!C^{{}}_{12} & C^{{}}_{11} & C^{{}}_{12} & 0 & \!\!0 & \!\!0 \!\!\\[1mm]
 \!C^{{}}_{12} & C^{{}}_{12} & C^{{}}_{11} & 0 & \!\!0 & \!\!0 \!\!\\[1mm] \hline
 \!0 & 0 & 0  & \!C^{{}}_{44} & \!0 & \!\!0 \!\!\\[1mm]
 \!0 & 0 & 0  & 0 & \!\!\!C^{{}}_{44} & \!\!0 \!\!\\[1mm]
 \!0 & 0 & 0  & 0 & 0 & \!\!\!C^{{}}_{44} \!\!\!
 \end{array} \right) \!\!
  \left( \begin{array}{c}
 \!\pp^{{}}_x u^{{}}_x \\[1mm]
 \!\pp^{{}}_y u^{{}}_y \\[1mm]
 \!\pp^{{}}_z u^{{}}_z \\[1mm] \hline
 \!\!\pp^{{}}_y u^{{}}_z\!+\!\pp^{{}}_z u^{{}}_y\!\! \\[1mm]
 \!\!\pp^{{}}_x u^{{}}_z\!+\!\pp^{{}}_z u^{{}}_x\!\! \\[1mm]
 \!\!\pp^{{}}_x u^{{}}_y\!+\!\pp^{{}}_y u^{{}}_x\!\! \end{array} \right).
 \eeq
Here, $C^{{}}_{ik}$ are the elastic moduli which are listed for SLS float glass in \tabref{all_param}, and $C_{12} = C_{11} - 2C_{44}$.

In the piezoelectric AlN film, the stress tensor $\sigmabf$ and the electric displacement field $\DDD$ are given by the strain tensor $\frac12\big[\nablabf\uuu + (\nablabf\uuu)^T\big]$, the electrical field $-\nablabf\varphi$, the electromechanical coupling coefficients $e_{ik}$, the elastic moduli $C_{ik}$, and the electric permittivities $\varepsilon_{ik}$ as
 \bal
 &\left( \begin{array}{c}
 \sigma^{{}}_{xx} \\  \sigma^{{}}_{yy} \\  \sigma^{{}}_{zz} \\ \hline
 \sigma^{{}}_{yz} \\  \sigma^{{}}_{xz} \\  \sigma^{{}}_{xy} \\ \hline
 D^{{}}_x \\ D^{{}}_y \\ D^{{}}_z \\
 \end{array}  \right)
 =
 \eqlab{piezoEqu}
 \\ \nn
 &\left( \begin{array}{c@{\:}c@{\:}c@{\:}|c@{\:}c@{\:}c@{\:}|c@{\:}c@{\:}c}
 \!\!C^{{}}_{11} & C^{{}}_{12} & C^{{}}_{13} & 0 & 0 & 0 & 0 & \!0 & \!\text{-}e^{{}}_{31} \!\!\\
 \!\!C^{{}}_{12} & C^{{}}_{11} & C^{{}}_{13} & 0 & 0 & 0 & 0 & \!0 & \!\text{-}e^{{}}_{31} \!\!\\
 \!\!C^{{}}_{13} & C^{{}}_{13} & C^{{}}_{33} & 0 & 0 & 0 & 0 & \!0 & \!\text{-}e^{{}}_{33} \!\!\\ \hline
 \!\!0 & 0 & 0 & \!C^{{}}_{44} & 0 & 0 & 0 &  \!\text{-}e^{{}}_{15} & \!0 \!\!\\
 \!\!0 & 0 & 0 & 0 & \!C^{{}}_{44} & 0 & \text{-}e^{{}}_{15} & \!0 & \!0    \!\!\\
 \!\!0 & 0 & 0 & 0 & 0 & \!C^{{}}_{66}  & 0 & \!0 & \!0                     \!\!\\ \hline
 \!\!0 & 0 & 0 & 0 &  \!e^{{}}_{15} & 0 & \ve^{{}}_{11} &  \!0 & \!0        \!\!\\
 \!\!0 & 0 & 0 &  e^{{}}_{15} & 0 & 0 & 0 & \!\ve^{{}}_{11} & \!0         \!\!\\
 e^{{}}_{31} & e^{{}}_{31} & e^{{}}_{33} & 0 & 0 & 0  & 0 & \!0 & \!\ve^{{}}_{33}\!\!\\
 \end{array}  \right) \!\!
 \left(   \begin{array}{c}
 \pp^{{}}_x u^{{}}_x \\ \pp^{{}}_y u^{{}}_y \\ \pp^{{}}_z u^{{}}_z \\ \hline
 \!\!\!\pp^{{}}_y u^{{}}_z \!+\!  \pp^{{}}_z u^{{}}_y \!\!\!\\
 \!\!\!\pp^{{}}_x u^{{}}_z \!+\! \pp^{{}}_z u^{{}}_x  \!\!\!\\
 \!\!\!\pp^{{}}_x u^{{}}_y \!+\!  \pp^{{}}_y u^{{}}_x \!\!\!\\ \hline
 -\pp^{{}}_x \varphi \\ -\pp^{{}}_y \varphi \\ -\pp^{{}}_z \varphi \\
 \end{array}\right)\!.
 \eal
Here, the symmetry of AlN leads to $C_{12} = C_{11} - 2C_{66}$.

In terms of volume (mass) ratios, the device consists of 99.8\% (99.5\%) glass, 0.2\% (0.3\%) AlN film, and 0.0\% (0.2\%) metal electrodes. We have found that within the numerical uncertainties, the simulation results are unchanged when leaving out the metal electrodes. So to reduce the memory requirements and to speed up computation times, we have left out the tiny metal electrodes in our modeling. Consequently, the boundary conditions for the electric potential $\varphi$ are: (1) on the interface between the bottom electrode and the AlN film, the potential is grounded, $\varphi = 0$; (2) on the interface between one half of the split electrode and the AlN film, $\varphi = +\frac12 V_0$; (3) on the interface between the other half of the split electrode and the AlN film it is in anti-phase, $\varphi = -\frac12 V_0$; (4) on the sides walls with normal vector $\nnn$ of the AlN film, the no-free-charge condition applies, $\nnn\cdot\nablabf\varphi = 0$. For the mechanical displacement filed in the solids, the non-stress condition applied to all external surfaces, $\nnn \cdot \sigmabf = \zerovec$.

\begin{table}[t]
\centering
\vspace*{-2.5mm}
\caption{\tablab{all_param} Material parameters at $25~\SICel$ for the AlN-thin-film transducer and the SLS float-glass substrate. The frequency-dependent damping coefficients $\Gamsl(f)$ and $\Gamma_\ve(f)$ are obtained by fitting to the measured width of the resonance peaks. The tiny metal electrodes are not simulated in this study.}
\begin{ruledtabular}
\begin{tabular}{ccccc}
 Parameter &  Value & & Parameter  & Value \\ \hline
 \multicolumn{5}{l}{\emph{Thin film aluminum nitride}, AlN \cite{Caro2015}}  \upspace \downspace \\
 $\rho $    & 3300 $\SIkg\:\SIm^{-3}$ &   & $\Gamsl$ & 0.001 \\
 $C_{11}$   & 410.2 GPa  &  & $C_{33}$	& 385.0 GPa\\
 $C_{12}$   & 142.4 GPa  &  & $C_{44}$	& 122.9 GPa\\
 $C_{13}$   & 110.1 GPa  &  & $C_{66}$	& 133.9 GPa\\
 $e_{13}$   &$-1.0475~\SIC\:\SIm^{-2}$    &
 & $e_{15}$	& $-0.39~\SIC\:\SIm^{-2}$ \\
 $e_{33}$   & 1.46 $\SIC\:\SIm^{-2}$      &
 &  $\Gamma_\ve$ & $-\frac{1.43\,\SIMHz + 0.98 f}{1\,\SIGHz}$ \\
 $\epsilon_{11}$   &  9.21 $\epsilon_{0}$ &
 & $\epsilon_{33}$ & 10.12 $\epsilon_{0}$
 \\[2mm]
 \multicolumn{5}{l}{\emph{SLS float-glass substrate}  \cite{Schott_float_glass}} \\
 $\rho$   & (2522 $\pm$ 15) $\SIkg\:\SIm^{-3}$ &  \multicolumn{3}{c}{(measured at EPFL) \upspace \downspace}\\
 $E$      & $(72\pm2)$ GPa        &  & $s$  & $0.23\pm0.01$           \\
 $C_{11}$ & 83.5 GPa      &  & $C_{44}$    & 29.3 GPa        \\
 $C_{12}$ & 24.9 GPa      &  & $\Gamsl(f)$  & $f^{-1}$ 0.018~MHz  \\
 $\cL$ & 6118 $\SImps$    &  & $\cT$    & 3623 $\SImps$    \\
\end{tabular}
\end{ruledtabular}
\end{table}

Once a solution to $\uuu$ and $\varphi$ has been obtained, the electrical impedance $Z$ can be computed in the model as, the ratio of the total voltage drop $V_0$ and the integral over the electrical current density $\JJJ$ through the surface $\pp\Omega_{\frac12}$ one half of the split electrode
 \beq{ZsimDef}
 Z = \frac{V_{0}}{I} = \frac{V_{0}}{\int_{\pp\Omega_{\frac12}}\JJJ\cdot\nnn\:\dm a}.
 \eeq
As there are no free charges in the system, the only contribution to the current density $\JJJ$ is the polarization current density $\nnn\cdot\JJJ = \eee_z\cdot\JJJ_\mr{pol} = \pp_t P_z = -\ii\omega(D_z + \varepsilon_0\pp_z\varphi)$. We compute $D_z$ from \eqref{piezoEqu} and obtain, as in Ref.~\cite{Skov2019b},
 \beq{Zsim}
 Z = \frac{\frac{\ii}{\omega} V_{0}}{\int_{\pp\Omega_{\frac12}}
 \!\!\big[e_{31}(\pp_xu_x\!+\!\pp_yu_y) \!+\! e_{33}\pp_z u_z
 \!+\! (\varepsilon_0\!-\!\varepsilon_{33})\pp_z\varphi\big]\dm a}.
 \eeq

We have implemented the model of the AlN-thin-film-actuated glass-block devices in the finite-element-method software Comsol Multiphysics 5.4 \cite{Comsol54} following the procedure described in Ref.~\cite{Skov2019}. The coupled field equations~\eqnoref{EquMotionSolid} for the elastic-solid displacement $\uuu$ and the electric potential $\varphi$ are implemented together with the constitutive equations \eqsnoref{StressStrainSolid}{piezoEqu} using the weak form interface ``\textit{PDE Weak Form}". A COMSOL script with a PDE-weak-form implementation of acoustofluidics is available as supplemental material in Ref.~\cite{Muller2015}.

The system geometry described in \secref{design} is meshed fairly coarse, because we study the low end of the spectrum, mainly $f < 3$~MHz, where the wavelengths are larger than 1.1~mm. The mesh of the glass block is a free tetrahedral mesh with a maximum element size 0.2~mm, whereas the AlN thin-film is modeled as a swept, structured mesh with 2 mesh layers in the thickness direction and the same element size along the surface as for the glass. We used quadratic Lagrangian test functions, and used the mesh convergence test described in Ref.~\cite{Muller2012}, to verify that the numerical resolution was satisfactory.

We used an adaptive step-size algorithm to compute the impedance spectra. Close to a resonance peak, the step size went down to 15~Hz, but far from resonance peaks, it increased to 1600~Hz. The resonance frequency was determined as the position of the maximum of a Lorentzian fit containing the 11 closest points. A typical small 0.1-3-MHz spectrum contained 450-500 points, whereas a long 0.1-10-MHz spectrum contained 2500-2800 points. A short and long spectrum took respectively around 0h:40min and 2h:50min to compute on our HP-G4 workstation with a 3.7-GHz Intel Core i9 7960X-2.8 16-core processor and with a memory of 128 GB RAM.

\begin{figure}[!b]
\centering
\includegraphics[width=\columnwidth]{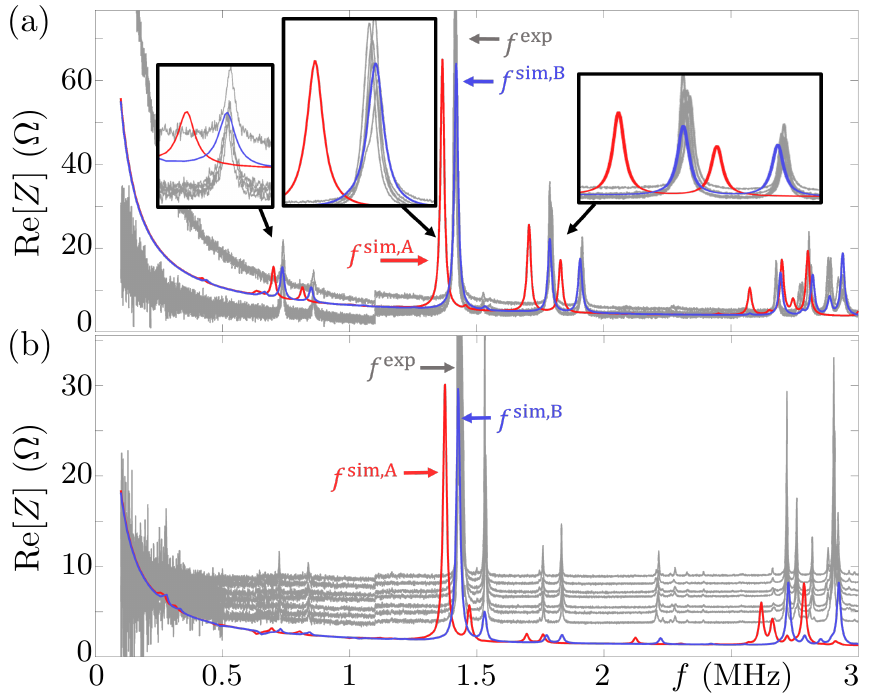}
\caption{\figlab{ReZExpSim} (a) The resistive real part $\re[Z]$ of the impedance as a function of frequency $f$ measured for the 5 devices C2 - G2 of \figref{Setup_and_measurements}(d) (gray curves $f^\mr{exp}$) and simulated numerically using the parameter values from the literature, listed in  \tabref{all_param}, (red curve $f^\mr{sim,A}$), and using the fitted parameter values of \tabref{glass_param_fitted} (blue curve $f^\mr{sim,B}$). (b) Similar plot of $\re[Z]$ versus $f$, but for the 7 devices A8 - G8 of \figref{Setup_and_measurements}(c).}
\end{figure}

\section{Comparing simulation with experimental results}
\seclab{results}

We use the numerical model of \secref{modeling} to simulate electrical impedance spectra $Z^\mr{sim}(f)$ through \eqref{Zsim}. In the following, we compare these simulation results with experimental spectra $Z^\mr{exp}(f)$ described in \secref{Zexp}.

Two examples of such a comparison in the frequency range from 0.1 to 3~MHz are seen in \figref{ReZExpSim}(a) and (b) for device geometry 2 and 8, respectively, where we are plotting the resistive part $\re[Z]$ of the electrical impedance as an alternative way to reveal the resonance peaks, instead of using the phase spectra $\theta(f)$ as \figref{Setup_and_measurements}(c) and (d). The single red curve in each plot is the simulation results obtained with the parameter values listed in \tabref{all_param}, whereas the family of gray curves is the experimental result for all available devices with the geometry. It is seen that the experimental impedance spectra are reproduced fairly well by the simulation regarding the position of the resonance peaks and their relative heights. We note, that similar to \figref{Setup_and_measurements}, the off-resonance level of the experimental curves (gray) for the real part $\re[Z]$ coincide for the devices C2-G2 in \figref{ReZExpSim}(a) that have been discharged, while they are displaced for the devices A8-G8 in \figref{ReZExpSim}(b) that have not been discharged. Moreover, we note that in \figref{ReZExpSim}(a), we have fitted the off-resonance level in the numerical model for $Z$ in \eqref{Zsim} using  $\Gamma_\ve(f)$ given in \tabref{all_param}.

The agreement between experiment and simulation is quantified in \tabref{FreqResultsExpSim}, where it is seen that for the 173 resonance peaks in our study, the relative deviation $\Delta^\mr{A}_{k,n}$ of each simulated resonance frequency $f^\mr{sim,A}_{k,n}$ deviates only a few percent from the experimental value $f^\mr{exp}_{k,n}$. The mean value and the standard deviation of the mean value of all relative deviations $\Delta^\mr{A}_{k,n}$ is $(-4.2\pm0.04)$\%, which is a fairly good result. However, the simulation results show a statistical significant systematic underestimate of the resonance frequency. This is also clear from the insets in \figref{ReZExpSim}(a) showing a zoom-in on four resonance peaks: In all cases, the measured (gray) peaks of the seven nominally identical devices fall on top of each other, with variations much smaller than the peak width, whereas in contrast, the simulated (red) peaks are shifted downward in frequency by a few peak widths.

This systematic shift of the simulated resonance peaks leads us to investigate, if a fitting of the two most significant but not so accurately determined material parameters, Young's modulus $E$ and Poisson's ratio $s$ of the SLS float-glass substrate, could lead to better numerical agreement with the experiment. Indeed, by using a simple stepping procedure, we find that such an improvement is obtained by the substitution
 \beq{E_and_s_fit}
 E = 72~\SIGPa \xrightarrow[\text{fit}]{} 75.5~\SIGPa, \qquad
 s = 0.23 \xrightarrow[\text{fit}]{} 0.189.
 \eeq 
This procedure extends the methods for obtaining the electromechanical coefficients of piezoelectric transducers \cite{Lahmer2008, Kiyono2016} to determining the elastic coefficients of any elastic solid driven by vanishingly small thin-film transducers. The parameter values obtained for the best fitting to the experimental data are summarized in \tabref{glass_param_fitted}, and the resonance frequencies obtained from the simulation based on these values are denoted $f^\mr{sim,B}_{k,n}$.

The improved agreement between simulation and experiment is seen qualitatively in \figref{ReZExpSim}, where the blue curves representing $f^\mr{sim,B}_{k,n}$ more accurately reproduce the gray experimental impedance spectra. The improved agreement is also evident from the insets in \figref{ReZExpSim}(a), where the blue curve $f^\mr{sim,B}_{k,n}$ now falls on top of the gray experimental curves within a fraction of the peak width. Quantitatively, the improved agreement between experiment and simulation is seen in \tabref{FreqResultsExpSim}, where the relative deviation $\Delta^\mr{B}_{k,n}$ of each simulated resonance frequency $f^\mr{sim,B}_{k,n}$ deviates less than one percent from the experimental value $f^\mr{exp}_{k,n}$. The mean value and the standard deviation of the mean value of all relative deviations is $\Delta^\mr{B}_{k,n} = (-0.03\pm0.04)$\%, or nearly 140 times better than the first result $(-4.2\pm0.04)$\%.

\begin{table}[t]
\centering
\vspace*{-2.5mm}
\caption{\tablab{glass_param_fitted} Fitted values for Young's modulus $E'$, Poisson's ratio $s'$, and the damping coefficient $\Gamsl$ of SLS float glass for improving the agreement between experiments and simulations. The values of  $C'_{11}$,  $C'_{12}$,  $C'_{44}$, $\cT'$, and $\cL'$ are functions of $E'$ and $s'$ \cite{Karlsen2015}.}
\begin{ruledtabular}
\begin{tabular}{lccccc}
 Parameter &  Value &$\qquad$& Parameter    & Value \\ \hline
 $E'$      & 75.5 GPa  & & $s'$         & 0.189   \upspace  \\
 $C'_{11}$ & 82.8 GPa      & & $C'_{44}$    & 31.8 GPa        \\
 $C'_{12}$ & 19.2 GPa      & & $\Gamsl(f)$  & $f^{-1}$ 0.018~MHz  \\
 $\cL'$    & 5731 $\SImps$ & & $\cT'$    & 3549 $\SImps$  \\
\end{tabular}
\end{ruledtabular}
\end{table}

The improved numerical agreement with experiments using the fitted values of Young's modulus and Poisson's ratio for the glass substrate suggests that our numerical fitting simulation algorithm may be used as an improved \emph{in-situ} determination of acoustic parameters in the 0.1-10~MHz range, and possibly for higher frequencies. At the present stage we do not know, if the relative deviation of roughly 4\% between the literature values in \tabref{all_param} and the fitted values in \tabref{glass_param_fitted} is within the uncertainty generated by sample-to-sample variations in SLS float glass, or if is a direct result of the fabrication process involving the deposition of the AlN-thin-film transducers at elevated temperatures. The improved simulation-to-experiment agreement also suggest that the impedance measurements coupled with simulations is a promising way to investigate the resonance modes, since there is little ambiguity in which peaks are which. Also from the measurements shown in \figref{ReZExpSim}, it is clear that by combining simulation with more varied device geometries it will be possible to predict particularly desired resonance frequencies with good accuracy and consequently to design appropriate device geometries.\\[-6mm]

\section{Conclusion}
\seclab{conclusion}

\vspace*{-3mm}
We have presented fabrication of 570-$\SImum$-thick, millimetric-sized SLS float-glass blocks with a 1-$\SImum$-thick AlN-thin-film piezoelectric transducer sandwiched between thin metallic electrodes and deposited on the top surface. The electromechanical properties of the devices were characterized by electrical impedance measurements in the frequency range from 0.1 to 10~MHz with a peak-to-peak voltage of 0.5~V applied to the electrodes.

We have measured the electrical impedance spectra of 35 different device geometries, all of width 2~mm, but with 9 different lengths ranging from 2 to 6~mm and with 2-7 copies of each individual geometry. Each impedance spectrum exhibits many resonance peaks, of which we carefully measured the 5 most prominent ones in each spectrum. Comparing the resulting 173 experimental resonance frequencies with the result of a finite-element method simulation, using material parameter values from the manufacturer, results in an average relative deviation of the 173 simulated resonance frequencies from the experimental ones of $(-4.2\pm0.04)$\%.

We have shown that by using Young's modulus and Poisson's ratio of the SLS float glass as fitting parameters in the simulation, we could reduce the relative deviation of the 173 simulated resonance frequencies from the experimental ones by a factor of nearly 140 to $(-0.03\pm0.04)$\%. This result suggests that our numerical fitting procedure could lead to an improved method for \emph{in-situ} determination of the acoustic parameters at ultrasound frequencies of any elastic solid onto which a thin-film transducer can be attached.

\section*{Acknowledgements}

This work was supported by the \textit{BioWings} project funded by the European Union's Horizon 2020 \textit{Future and Emerging Technologies} (FET) programme, grant No.~801267.


%
%


%

\end{document}